# QUANTUM COMPUTING IN INDUSTRIAL ENVIRONMENTS: WHERE DO WE STAND AND WHERE ARE WE HEADED?

Eneko Osaba[1], Iñigo Perez Delgado[2], Alejandro Mata Ali[2], Pablo Miranda-Rodríguez[1],

Aitor Moreno Fdez de Leceta[2], Luka Carmona Rivas[2]

[1] TECNALIA, Basque Research & Technology Alliance (BRTA), 48160 Derio, España

[2] Ibermatica Fundazioa, Parque Tecnológico de Bizkaia, Ibaizabal Bidea, Edif. 501-A, 48160 Derio, España

## Abstract

*This article explores the current state and future prospects of quantum computing in industrial environments. Firstly, it describes three main paradigms in this field of knowledge: gate-based quantum computers, quantum annealers, and tensor networks. The article also examines specific industrial applications, such as bin packing, job shop scheduling, and route planning for robots and vehicles. These applications demonstrate the potential of quantum computing to solve complex problems in the industry. The article concludes by presenting a vision of the directions the field will take in the coming years, also discussing the current limitations of quantum technology. Despite these limitations, quantum computing is emerging as a powerful tool to address industrial challenges in the future.*

## 1.- INTRODUCTION

The first half of the 20th century shook the foundations of human society. In the scientific and technological realm, two theories—General Relativity and Quantum Mechanics—conceptually revolutionized the old structures of physics. This revolution, as the meaning of the word demands, was not limited to the construction of a new worldview, but also brought with it significant technological advancements. In particular, quantum mechanics has been especially prolific as the foundation for a vast number of technological discoveries. The main fields impacted have been those that, by their nature, are governed by the laws of quantum theory; for example, the construction of microscopic objects, such as nanotechnology, as well as fields like electronics or nuclear medicine.

The scientific field in which quantum mechanics has generated the most expectations over the past two decades is computing. However, as early as the 1980s, Yuri Manin [1] and Richard Feynman [2] pointed out that the capabilities of a quantum computer—hypothetical at the time—would surpass those of any classical computer. The initial goal of this device was to simulate certain physical systems, given that traditional computers had (and still have) severe limitations when it comes to calculating the superpositions and entanglements characteristic of many-body quantum systems. This is due to the exponential amount of classical resources required to emulate increasingly large quantum systems. This issue disappears if qubits (quantum bits) are used as the basic units of computation. Qubits intrinsically obey the laws of quantum mechanics, so when working with them, it is no longer necessary to emulate superposition and entanglement. This was precisely Manin and Feynman's intuition: we know that a quantum system must be efficiently simulable by another quantum system since they share the same nature. If quantum systems cannot be efficiently simulated by classical computers, then, by necessity, quantum computers will have an advantage over traditional computers.

These seminal works from the 1980s laid the foundation for the field now known as quantum computing. Since then, researchers have developed a wide range of algorithms that extend beyond simulation into areas such as optimization, machine learning, and cryptography. Some of these algorithms represent both a quantitative and qualitative leap compared to existing classical algorithms.

However, contrary to an idea entrenched in the popular imagination—that quantum computers will one day replace classical computers—we want to emphasize that this idea is false. Quantum computers are special-purpose devices[1] and

---

[1] Special-purpose computers are devices systems designed to perform specific tasks or a limited set of tasks. They are distinguished from general purpose computers, which are designed to be extremely flexible and to perform multiple tasks (e.g. a personal computer).

| | Quantum Computing in Industrial Environments: where do we stand and where are we headed? | |
|---|---|---|
| | E. Osaba, I. Perez Delgado, A. Mata Ali, P. Miranda-Rodríguez, A. Moreno Fdez de Leceta, L. Carmona Rivas | |

will complement their classical counterparts. The quantum advantage stems from the ability of qubits to perform a broader set of operations compared to traditional bits. However, this increased versatility does not confer an advantage for all tasks [3, 4]. Combined with the high cost of building a quantum computer, this means that the vast majority of tasks will continue to be performed on traditional computers (such as adding two real numbers or opening a word processor). Moreover, there are no quantum algorithms (nor are any expected) that computationally optimize these kinds of tasks.

Today, quantum computers are already a reality. There is a genuine ecosystem of quantum computers based on different hardware technologies, each with its own advantages and disadvantages. Some examples include superconducting chips (IBM, Google, D-Wave, etc.), trapped-ion systems (IonQ, Quantinuum, etc.), photonic systems (Xanadu, PsiQuantum, etc.), and neutral atom systems (Pasqal, QuEra, etc.). This technological diversity contrasts with one consistent feature of these systems: the number of logical qubits[2] in most of them ranges from several dozen to around a hundred. In this context, D-Wave computers are an exception, as their unique characteristics allow them to have thousands of logical qubits, as we will explain in Section 2.2.

In addition to the limitation on the number of qubits, it is important to consider the issues of noise and qubit coherence time. Qubits are extremely sensitive to interactions with their environment (noise), which can alter their states to undesired configurations [5]. Moreover, qubits exhibit their quantum properties for a limited period, known as coherence time, after which quantum computation can no longer be performed with them [6]. All of this restricts the current range of applications for quantum computers, especially in industrial settings. For this reason, the scientific and technical community considers the field to be in the so-called NISQ era (Noisy Intermediate-Scale Quantum [7]), where there is not yet a true quantum advantage—that is, quantum computers still cannot outperform the capabilities of current classical supercomputers. That does not mean, as we will see in Section 3, that NISQ computers are not capable of solving problems of industrial interest.

## 2.- COMPUTATIONAL PARADIGMS

In the realm of quantum computing, two paradigms are most commonly cited: gate-based quantum devices and quantum annealing computers. Additionally, a third paradigm can be considered, Tensor Networks, which are part of the category known as quantum-inspired algorithms. Here is a brief overview of each of these three concepts.

### 2.1.- Gate-based quantum computers

Gate-based quantum computers are the most similar to current general-purpose classical computers. In fact, they are a generalization of these: they can, in principle, perform any classical computation, using qubits as bits, that is, moving them only from state 0 to state 1 and back. However, the true power of gate-based quantum computers lies in their ability to control qubits beyond the binary: if we represented all the possible states of a qubit on a sphere (as shown in Figure 1), the classical states would be only the north pole (state 0) and the south pole (state 1). In this analogy, the rest of the "planet's" surface would represent places accessible only to the qubit.

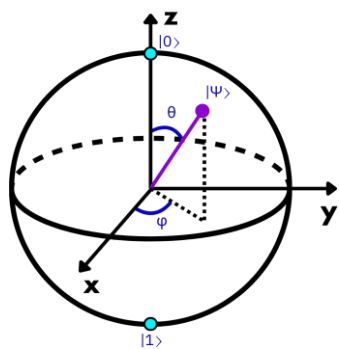

*Figure 1: Graphic representation of all possible states of a qubit known as the "Bloch Sphere".*

---

[2] Logical cubits refer to the "effective" cubits, i.e. the cubits that the computer can use for its calculations. To create a logical cubit, several physical cubits are needed, which are the "real" cubits, made with hardware. Physical cubits can be used, among other things, to correct errors or to increase the connectivity of a logical cubit to other logical cubits.

Quantum gate computers, being generalizations of classical general-purpose computers, are useful for a wide range of tasks. The most well-known quantum algorithms are Grover's algorithm [8], which is capable of parallelizing the search for objects in lists, and Shor's algorithm [9], which takes advantage of a mathematical technique called the "Chinese Remainder Theorem" to find the two prime factors of a number constructed as the multiplication of those numbers. Being able to find these two prime factors renders RSA-based encryption, one of the most widely used today, obsolete. Therefore, there is currently an effort to find and implement new classical encryption paradigms that take into account the capabilities of quantum computing: "post-quantum" or "quantum-safe" paradigms.

These two algorithms, Grover's and Shor's, were developed in the 90s and are not particularly suitable for the NISQ era due to the number of gates they require, their sensitivity to noise, and the coherence limitations in current devices. However, there are multiple algorithms designed for implementation in current NISQ computers. Some of them are optimization algorithms, such as the variational algorithms QAOA (Quantum Approximate Optimization Algorithm, [10]) and VQE (Variational Quantum Eigensolver, [11]), and others are specific mathematical operations designed as part of more general methods, such as QFT (Quantum Fourier Transform, [12]) or QPE (Quantum Phase Estimation, [13]). There is also work focused on QML (Quantum Machine Learning, [14]), which reports promising accuracy and sensitivity compared to their classical counterparts, even with current NISQ computers.

In the same way that in the 1940s the ENIAC computer used thermionic valves and current classical computers use silicon chips, there exist multiple hardware technologies that could be used as physical qubits. One of the most promising is quantum gate computers based on superconducting circuits (such as those built by IBM, Rigetti, IQM...), which exploit the quantum properties of superconducting currents. The biggest issue with these computers is the cryogenics required to achieve said superconductivity, which is one of the reasons behind the famous quest for room-temperature superconductors.

On the other hand, as mentioned in the introduction, we have the technology of electromagnetically trapped ions (IonQ, Quantinuum, Honeywell...), where quantum information is encoded in the electronic state of these ionized atoms. In photonic computing (PsiQuantum, Photonic, AegiQ...) the polarization and color of photons are taken into account to transmit quantum states. Cold neutral atom computers (Atom Computing, Pasqal, ColdQuanta...) encode information in atoms cooled by laser pulses to the point where they begin to exhibit quantum properties that allow their use as qubits. Nitrogen vacancy computers, or NV for short (Quantum Brilliance, Turing, Xeeq...), use electron holes in diamonds doped with this element as highly stable qubits. There are also solid defect technologies based on other materials, such as boron, silicon carbide, or tin. Quantum dots (Equal1, Intel, QuantumMotion...) are another candidate technology for creating qubits, consisting of nanocrystal semiconductor systems that exhibit quantum properties in which the qubit is encoded. Finally, there are topological qubits (Microsoft, Nokia, Quoherent...), which, although still in a very experimental stage, promise superior stability as they can avoid cumulative errors by storing the quantum state in the topological properties of the material.

## 2.2.- Quantum Annealers

Quantum annealers are specialized quantum devices intended to solve combinatorial optimization problems. They employ a heuristic algorithm similar to simulated annealing in classical computing. By utilizing quantum mechanics, these computers efficiently navigate the solution space of a problem to identify the lowest energy state.

The quantum annealing process begins by representing the problem as an energy landscape. Initially, the quantum system is prepared in a superposition of all possible solutions, representing a high-energy state. As the system evolves, the Hamiltonian is adjusted, allowing the system to explore the energy landscape. The goal is to guide the process towards the minimum energy state, which corresponds to the optimal solution of the problem.

Unlike gate-based quantum devices that execute calculations through a series of gates, quantum annealers function by continuously evolving an Ising Hamiltonian under the guidance of the Schrödinger equation. While this approach is particularly suitable for solving certain problems, it lacks the inherent universality of gate-based systems.

Progress in quantum technologies has led to the development of intermediate-scale quantum annealers designed for programmable use. In this context, the Canadian company D-Wave has developed commercial devices based on quantum annealing, enhancing the accessibility of this technology. Currently, D-Wave's most advanced computer, called Advantage_system6.4, consists of 5,612 superconducting qubits, arranged in a Pegasus topology [15].

Finally, it is important to highlight that D-Wave devices are currently the most widely used for optimization problems approached from a quantum perspective. Interesting research has been published in fields such as finance [16], logistics [17], and industry [18].

## 2.3. - Tensor networks

Tensor networks (TN) are a type of computation usually called quantum-inspired, which generalizes quantum computation and allows the simulation of quantum systems and algorithms of certain characteristics, as well as more general processes. The basis of TNs is the graphical representation of a tensor algebra equation, allowing the handling of extremely complex expressions in a simple way, enabling the creation of algorithms with the same [19, 20]. This property allows us to do two very useful tasks: compress a dataset and create a dataset. It is important to note that this technology is not purely quantum, but is a quantum-inspired classical technology, since it runs on classical devices.

The first of these tasks is to take a certain tensor representing a dataset and transform it to a tensor network representation that, when operated on, returns the same element. This, in certain occasions, will lead us to the fact that the TN representing that object will have a lower number of elements than the represented tensor. That is, it will occupy less memory. An example of three representation schemes can be seen in Figure 2. We can perform this compression in an exact or approximate way, by preserving the underlying structures that generate such a tensor based on certain relations and logics. In this way, we can efficiently represent quantum states of many qubits, such as giant layers of deep learning models or big data datasets. This allows us to handle a larger amount of information, analyze it more easily and simulate larger systems.

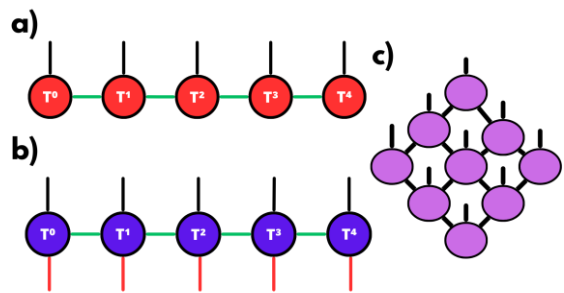

*Figure 2: Main types of tensor network representations. a) Matrix Product State (MPS), b) Matrix Product Operator (MPO), c) Projected Entangled-Pair States (PEPS).*

The second of these tasks is the generation of data tensors that follow certain logics. That is, it allows us to generate a dataset from a TN that has underlying structures that implement some properties or characteristics of a problem, the final tensor being the solution of the problem itself. To do this, each tensor of the network interacts with those to which it is connected, as a set of individuals cooperating to carry out a task. Its clearest application is combinatorial optimization, since the tensors can emulate the superposition of all possible outcomes and evaluate them, discarding non-viable configurations and minimizing those that are feasible. An example of this type of network is shown in Figure 3, which is used to solve the traveling salesman problem.

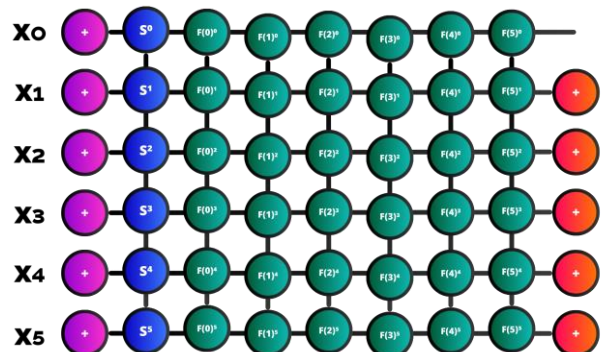

*Figure 3: Tensor network solving the traveling salesman problem [26]. Read from left to right, this network starts by considering all possible combinations with the purple "+" nodes, associates to each one a value that will be smaller the higher the cost of that path with the blue "S" nodes, and finally discards the paths in which cities are repeated with the green "F" nodes. With this state created, we determine the position of its largest element by the partial sum of all elements with the same first variable, using the red "+" nodes in all variables except the one to be determined.*

Both capabilities have been used in the state of the art to compress large language models, or LLMs [21], allowing them to occupy 93% less memory and to be trained twice as fast. TNs have also been applied to materials simulation [22], or to solve problems such as portfolio optimization [23] or other combinatorial optimization problems [24, 25, 26].

### 3.- QUANTUM COMPUTING APPLIED TO INDUSTRIAL ENVIRONMENTS

The nascent stage of quantum computers has not discouraged the scientific community from proposing resolution methods that can genuinely add value in specific real-world applications. In pursuit of such techniques, researchers have concentrated on designing and implementing hybrid methods that seek to leverage the advantages of both computational paradigms (classical and quantum). As a result, a notable portion of the scientific community is devoted to developing these types of approaches.

Regarding industrial problems, one example that has been frequently addressed from the perspective of quantum computing is the Bin Packing Problem. This problem, illustrated in Figure 4, focuses on optimizing the packing of products into a finite number of containers, which is a daily and essential task in the field of production and distribution. Additionally, depending on the characteristics of both the packages and the containers, multiple variants of the problem can be formulated. In [27], for instance, a hybrid algorithm utilizing D-Wave annealers is introduced to address a three-dimensional bin packing problem. This algorithm takes into account constraints such as package fragility, compatibility and incompatibility of various types of goods, and the presence of centers of gravity within the container. In [28], the same authors proposed a variant of this system, specifically aimed at packing goods in logistics trucks. This study incorporates additional constraints, such as the delivery sequence of the packages, ensuring that the packages to be delivered first are positioned closest to the vehicle's door.

The utility of the bin packing problem is also evident in works such as [29] and [30]. In the former, an issue related to the optimal loading of a cargo plane is addressed, while in the latter, a crucial problem in the atomic energy industry is tackled. This problem involves the filling of spent nuclear fuel into deep storage containers.

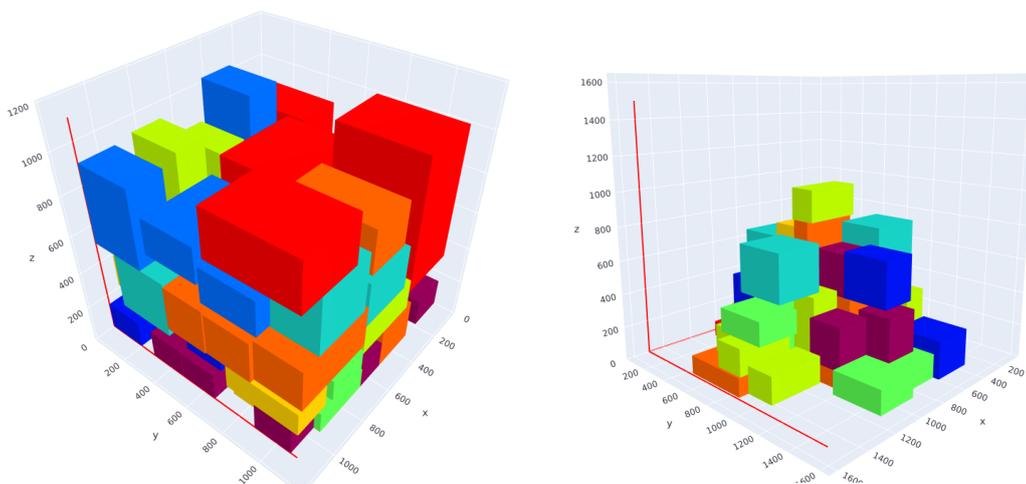

*Figure 4: Two examples of solutions to the three-dimensional bin packing problem. Images extracted from [27]*

Another problem widely addressed from the quantum perspective is the Job-Shop Scheduling Problem (JSSP). This problem, illustrated in Figure 5, involves planning the production of a series of jobs on a specific set of machines. Each job consists of a number of tasks that must be executed in a specific order. Moreover, each task has a fixed duration and can only be executed on one of the available machines. Finally, it is important to note that each machine can only perform one task at a time. The goal of the JSSP is to minimize the total production time.

This problem, which has significant applications in industrial environments, has been the subject of numerous studies focused on different quantum paradigms. For instance, [31] presents an implementation of a Quantum Approximate Optimization Algorithm (QAOA) to solve this problem, while [32] describes an implementation based on quantum annealing, executed on devices provided by D-Wave.

Beyond these studies, several variants of the JSSP have been tackled using quantum computing. For instance, [33] explores a flexible version of the problem where job processing is not limited to a single machine. The same authors introduce an interesting variation in [34], which accounts for dynamic execution times. In both scenarios, the quantum annealer paradigm is employed to solve the problems. Additionally, in the context of TNs, the research in [25] is noteworthy for addressing a variant known as task optimization.

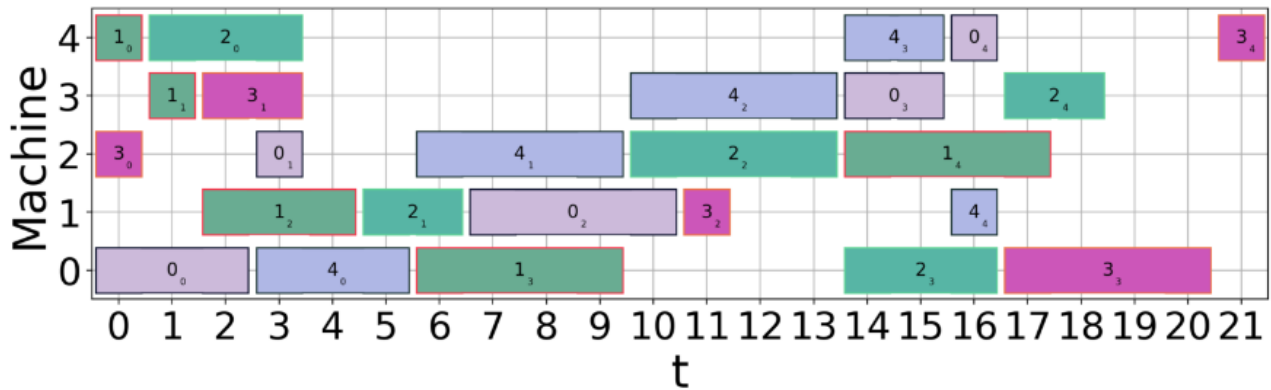

*Figure 5: Solution to an example instance of the JSSP consisting of 5 jobs, each defined by 5 tasks with durations ranging from 1 to 4 time pulses. In the image, each task is indicated by the job number (textual and color-coded) and the task number (subscript). Since there are 5 tasks per job, 5 different machines are also required, arranged on the vertical axis. This image was generated by the website described in Section 4.*

If we turn our attention to manufacturing-related problems, it is interesting to highlight the research proposed in [35], which presents a hybrid technique for route planning in multi-robot environments. This technique allows robotic units to carry out assigned tasks without colliding with each other during the process. In the article, researchers explore the effectiveness of a hybrid approach where the D-Wave quantum computer solves different instances of the problem, while various classical mechanisms automatically analyze the solutions to accept, reject, or even correct them. In connection with this topic, the work presented in [36] by researchers from the automotive brand BMW is noteworthy. It focuses on planning the routes that robots must follow to carry out their tasks within a vehicle assembly plant

Another field where quantum computing has shown its potential is logistics. In this context, it is interesting to highlight the work proposed in [37], which presents a hybrid technique for solving the vehicle routing problem. In this study, the authors divide the problem into two completely distinct phases. In the first phase, they assign customers to different routes, thus creating an association between customer and route, and generating a set of subproblems. This first process is carried out using classical clustering algorithms. Once this step is completed, the different subproblems are solved with the D-Wave quantum annealer, treating each of them as separate instances of the traveling salesman problem.

Adopting a more practical approach, the work presented in [38] addresses a route planning problem for heterogeneous trucks. Unable to solve the entire problem with a fully integrated quantum method, the authors propose an algorithm based on quantum annealing that iteratively assigns routes to trucks. This approach accommodates constraints such as pickup or delivery and time windows. Similarly, in [39], a goods distribution problem involving a heterogeneous fleet is tackled, allowing for the categorization of certain customers as priority, requiring service before a specific time threshold. Additionally, in the problem described in [39], some customers may have multiple orders, which can be fulfilled by more than one vehicle.

Finally, it is interesting to note that, beyond solutions based on quantum annealers, various studies have been recently published in the context of logistical problems, presenting other methods such as TNs or variational algorithms [26,40].

As might be expected, the number of works that hybridize quantum computing with industrial problems is much greater than those presented here. Throughout this section, several papers focused on very specific topics have been briefly described, with the intention of giving the reader an idea of the potential of this specific branch of research. Given the abundant associated research activity, it is prudent to conclude that quantum computing is emerging as a viable alternative to tackle complex real-world problems, at least on a proof-of-concept scale.

## 4.- DEMONSTRATOR APPLIED TO THE JOB SHOP SCHEDULING PROBLEM

Although significant progress has been made in democratizing quantum computing and making it freely accessible, this technology is still largely unfamiliar to the general public. Outside of specialized circles, it is rare to find researchers who have experimented with quantum algorithms, regardless of their simplicity. This undoubtedly poses a challenge to the development and popularization of the field, even within the scientific community.

Motivated by this situation and in alignment with the works highlighted throughout this article, the authors of this study have developed an online demonstrator for the previously described JSSP problem. This tool, available for free at https://www.q4real.eu/jssp/, enables any user, whether an expert or not, to perform small-scale experiments and access a significant variety of resolution methods.

Figure 7 displays the main page of the demonstrator. The central section of this page begins with a description of the JSSP problem. Following this, a brief tutorial is provided to help users navigate the tool. Lastly, this section details each of the resolution methods available in the current version of the demonstrator. Notably, it includes a diverse range of methods, some of which operate on various simulators, while others run on real quantum hardware, such as those utilizing D-Wave technology. Included among the methods are the CQM and BQM hybrid solvers from D-Wave, two TN-based methods, and an adaptive algorithm for solving QUBO format problems. Additionally, various combinations of these methods have been implemented. Detailed descriptions of each algorithm and their functionalities are available in the demonstrator. Interested readers are encouraged to visit the provided link for more information about the developed platform.

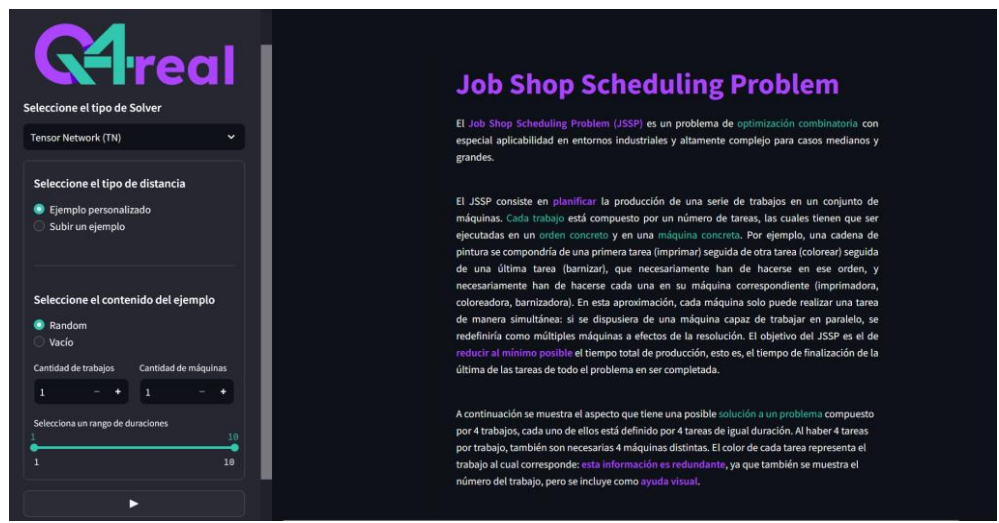

*Figure 7: Main page of the demonstrator developed for solving the JSSP problem.*

## 5.- WHERE IS THE FIELD HEADING?

We conclude this paper with some considerations about the future of the field. One of the main limiting factors of this quantum technology, aside from additional challenges, is the number of logical qubits. Therefore, we analyze below the temporal evolution of the number of qubits in IBM's machines, one of the leading providers of quantum computers. The strategy published in 2020 [41] was based on doubling or tripling the number of available qubits in quantum processing units (QPUs) each year, aiming to reach the range of 10,000–100,000 qubits starting in 2026. The development of QPUs aligned with these goals between 2016 and 2023.

Starting in 2023, following the presentation of its 1,121-qubit Condor chip [42], IBM proposed a shift in its technological strategy, resulting in a deviation from the previously established forecasts. This strategic shift involved a realignment of priorities, where the growth in the number of qubits took a backseat, prioritizing instead achieving higher qubit quality and increasing the number of quantum gates supported by the chips. Additionally, instead of "compact" QPU designs with many qubits, the current focus has shifted to the modular coupling of smaller chips. For example, systems with 399 qubits can now be reproduced using three Heron chips.

In any case, projections for the maximum number of next-generation qubits (through modular coupling) have been moderated to 1,092 qubits by 2028. Starting that year, new chips—Starling and Flamingo—are expected to be developed, featuring 200 and 2,000 qubits, respectively. IBM's current roadmap is represented graphically in Figure 7.

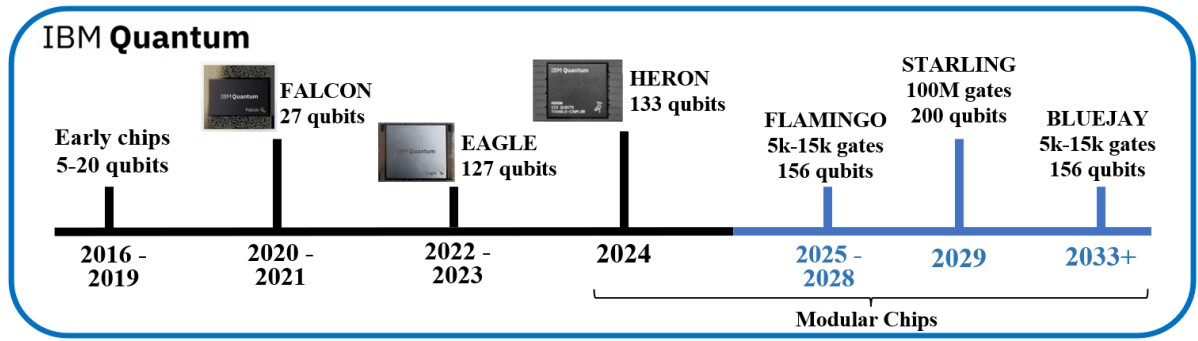

*Figure 7: IBM's roadmap for the development of QPUs (Quantum Processor Units). Designs from Heron onwards are modular [42].*

Shifting the focus, we now refer to another major player in the construction of quantum computers: D-Wave Systems. This company, founded in 1999 in the Canadian city of Burnaby, unveiled its first prototype in 2007, called ORION, which was composed of only 16 qubits. It wasn't until 2013 that D-Wave presented its first commercial computer with 128 qubits, named ONE, marking the beginning of a continuous path toward creating machines with greater capacity and better connectivity. We review this trajectory in Figure 8, where we can see how the latest commercial computer, dating from 2020, has the features mentioned in Section 2. In this image, we can discern another milestone in D-Wave's roadmap: the launch of the LEAP service. This was a breakthrough in the field, as LEAP offers a quantum cloud service, providing real-time access to the company's quantum computers as well as a portfolio of powerful hybrid solvers. In addition, D-Wave plans to release a new system in 2025, called Advantage2, which will feature 7,430 qubits arranged in a Zephyr topology [43].

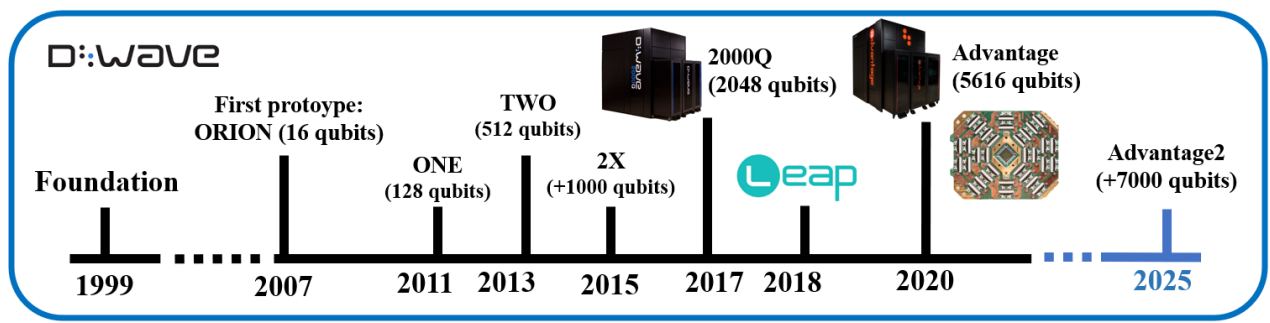

*Figure 8: D-Wave Systems' roadmap from its founding to the new Advantage2, expected in 2025.*

As previously outlined, there are other limiting factors in the context of quantum computing beyond the number of qubits that make up the systems. One of these limitations is related to the inherent error in qubits, due to factors such as decoherence or the difficulty of maintaining the strict environmental conditions (such as temperature). As a result, achieving reliable qubits to work with has become an absolute priority, leading companies like IBM to modify their roadmap. In recent years, significant attention has been given to the development of various error correction and mitigation protocols [44], as well as the creation of solution refinement mechanisms, such as the strategy developed by D-Wave known as reverse annealing [45].

Shifting gears, despite the nascent state of the field and the limitations that the main system manufacturers currently face, the interest and potential of the field are undeniable. This can be verified by the notable investments being made by various players and countries around the world. It is a fact that interest in quantum computing has grown significantly in recent years, both from the public and private sectors. In Spain, this has materialized in different projects and consortia,

| | Quantum Computing in Industrial Environments: where do we stand and where are we headed? | |
|---|---|---|
| | E. Osaba, I. Perez Delgado, A. Mata Ali, P. Miranda-Rodríguez, A. Moreno Fdez de Leceta, L. Carmona Rivas | |

such as Quantum Spain (with 22 million euros in public funds) [46], CUCO [47], and Q4Real [48]. The objectives of these initiatives are diverse, ranging from the application of quantum computing in industrial settings to the development and installation of quantum emulators and computers in various parts of the country. During this period, Spain has also seen a boom in the creation of companies dedicated to quantum computing, both in hardware and software, as well as the creation of quantum computing departments in large tech companies. Two companies that are paradigmatic of this trend are Qilimanjaro Quantum Tech and Multiverse Computing.

Outside of Europe, the most important players are China, the United States, and Canada. While China is a pioneer in quantum communication systems and Quantum Key Distribution (in fact, it is the leading power in this area), it lags behind its competitors when it comes to quantum computing. Nevertheless, China continues to make significant investments in this technology, estimated at 15 billion dollars [52], far surpassing the nearly 4 billion dollars invested by the U.S. government. However, it is in the United States and Canada where the most important quantum computing hardware companies, IBM and D-Wave, are located, although it is true that private investment in North America dropped by 50% in 2023 [53]. This decline in venture capital is partly driven by rising interest rates, the recent surge in Artificial Intelligence solutions (where investors find greater profitability), and a moderation in the short- and medium-term expectations regarding the utility of quantum computers in real industrial settings. For this reason, some actors are referring to the "quantum winter," a period of low investment and slowed development similar to those experienced by other technologies (for example, ZapataAI, one of the first quantum software startups, recently announced the cessation of operations [53]). However, it is important to emphasize that this budget reduction has not occurred in Europe, where private investment has actually increased by 3% [54].

Nonetheless, it is undeniable that quantum computing is a fascinating and constantly evolving field, promising true revolutions in its various application areas. The question is not whether these promises will come to fruition, but when.

| | Quantum Computing in Industrial Environments: where do we stand and where are we headed? | |
|---|---|---|
| | E. Osaba, I. Perez Delgado, A. Mata Ali, P. Miranda-Rodríguez, A. Moreno Fdez de Leceta, L. Carmona Rivas | |

## ACKNOWLEDGEMENTS

This work was supported by the Basque Government through HAZITEK program (Q4_Real project, ZE-2022/00033) and through ELKARTEK Program under Grants KK-2024/00105 (KUBIT project).